\documentclass[12pt]{article}
\usepackage[ansinew]{inputenc}

\usepackage{graphics}
\usepackage[dvipdfm]{graphicx}

\usepackage{epsfig,amssymb,graphicx}
\usepackage{natbib}

\usepackage{mathptmx}

\usepackage{setspace}
\author{T. C. O. Fonseca \footnote{(\baselineskip=10pt Department of Statistics,
Federal University of Rio de Janeiro and Statistics Department (University of Warwick, UK)}
\and V. S. Cerqueira 
\footnote{IPEA, RJ,  Brazil}
\and H. S. Migon \footnote{Department of Statistics,
Federal University of Rio de Janeiro, RJ, Brazil}
\and C. A. C. Torres \footnote{Universidad Nacional Autonoma de Honduras (UNAH)}
}

\title{The effects of degrees of freedom estimation in the Asymmetric GARCH model with Student-t Innovations}

\begin{document}

\def\spacingset#1{\renewcommand{\baselinestretch}%
{#1}\small\normalsize} \spacingset{2}

\maketitle


\begin{abstract}
This work investigates the effects of using the independent Jeffreys prior for the degrees of freedom parameter of a Student-t model in the asymmetric generalised autoregressive conditional heteroskedasticity (GARCH) model. To capture asymmetry in the reaction to past shocks, smooth transition models are assumed for the variance. We adopt the fully Bayesian approach for inference, prediction and model selection We discuss problems related to the estimation of degrees of freedom in the Student-t model and propose a solution based on independent Jeffreys priors which correct problems in the likelihood function. A simulated study is presented to investigate how the estimation of model parameters in the Student-t GARCH model are affected by small sample sizes, prior distributions and misspecification regarding the sampling distribution.  An application to the Dow Jones stock market data illustrates the usefulness of the asymmetric GARCH model with Student-t errors. \\

keywords: Heavy tailed distributions; Bayesian inference; Ill behaved likelihoods. 
\end{abstract}

\newpage
\spacingset{1.7} 

\section{Introduction}

This paper considers modelling future volatility using a generalised autoregressive conditional heteroskedasticity (GARCH) model on the past error terms and volatilities. The GARCH approach \citep{boll86} models the variance as a function of past values and an error term assuming that the variance is independent of shocks in the mean. The main advantage of these models is that they allow for accessing directly the effect of mean changes in the dynamics of the conditional variance. This is an important issue in financial time series, as changes in the mean tend to have a relevant impact on the uncertainty of the process under study. Several authors have considered GARCH models for financial time series. In particular, from a Bayesian perspective, \citet{delap00} proposed a full Bayesian analysis for GARCH models with Gaussian errors. An alternative approach is presented in \citet{Jac04} known as stochastic volatility modelling which allows for correlated conditional variance and mean. However, it is difficult to directly evaluate the past effects of the mean in the conditional variance. 

A stylized fact that needs to be captured by the models is that the conditional variance can react asymmetrically to positive versus negative shocks or large versus small shocks. That is, the conditional variance may follow different regimes according to the size and signal of the shock. News has an asymmetric impact on the economy and, for example, a large negative return might affect future volatility in a different way when compared to a positive return with the same size. \citet{Engle93} presents a review regarding this issue. \citet{Awar05} discuss the importance of asymmetries in the prediction of an economic index. This may be accommodated by smooth transition models based on an asymmetric specification of the conditional variance model. Thus, different specifications of the skedastic functions will take into account size and sign effects in the volatility. In this work, we consider smooth transition GARCH models and discuss inferential issues related to the smooth transition function and Bayesian inference.

\begin{figure}[htb] 
\begin{center} 
\includegraphics*[height=9cm,width=12cm]{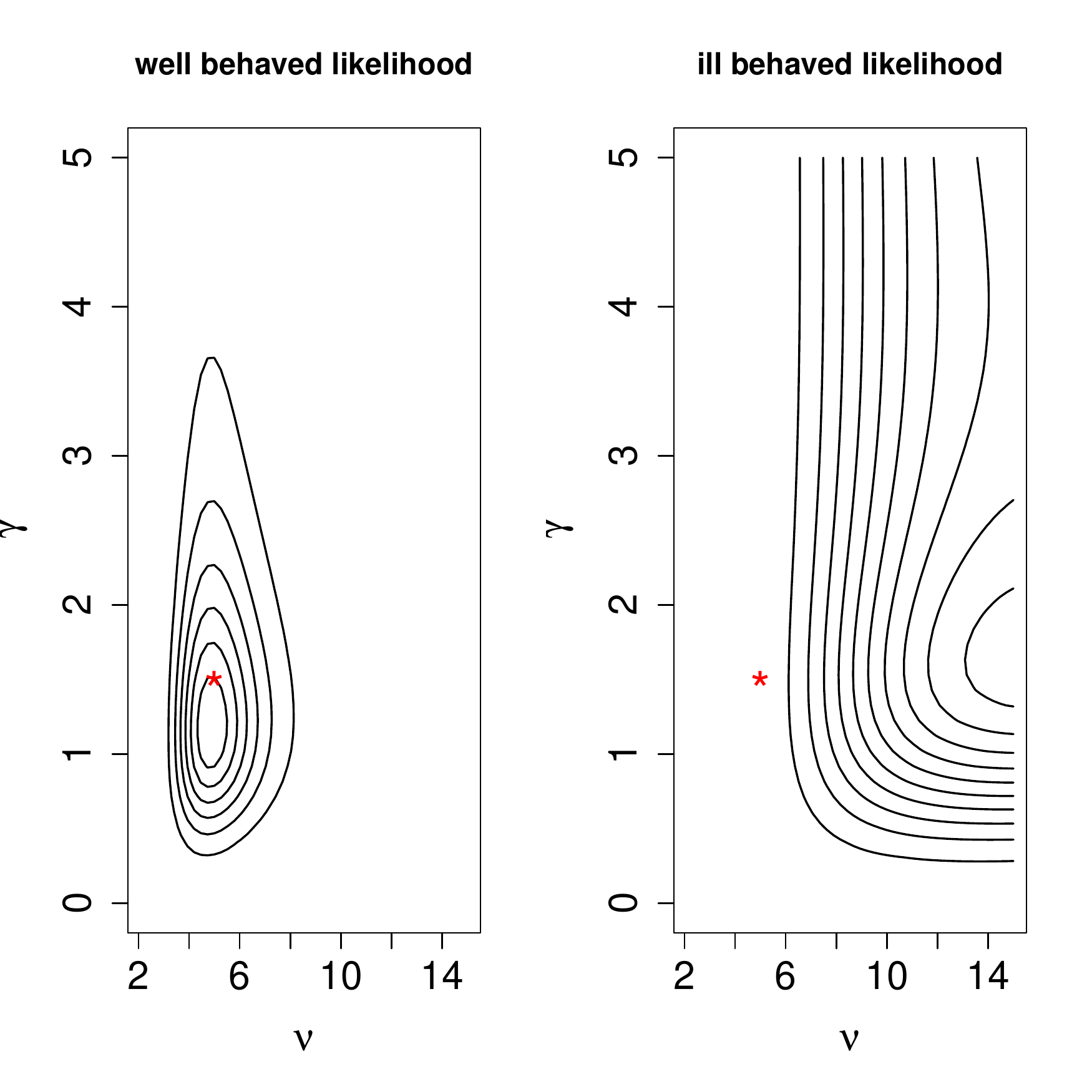} 
\caption{Contour plots of the joint likelihood function for $(\gamma,\nu)$ for two illustrative datasets of size 150. The red ``*" represents the true value of parameters. The parameters $\gamma$ and $\nu$ are the smooth transition and degrees of freedom parameters, respectively.}
\label{contour1}
\end{center}
\end{figure}

Regarding the sampling distribution for the error term, shocks are usually modelled as Gaussian distributed due mainly to mathematical convenience rather than being suitable for financial data. It is well known that financial time series exhibit heavier tails than allowed by the usual Gaussian model \citep{boll92}. In this work, we relax the assumption of Gaussian errors and consider Student-t distributions for the error terms in the GARCH model.  Since \citet{Man63} several authors have discussed the issue of fat tails in return datasets. \citet{boll87} introduced the GARCH-t model as a solution to the typical heavy tails of returns. Also in the context of robust analysis, \citet{Harvey08} proposed a Beta-t-EGARCH in which the volatilities depend on the score of a t distribution. \citet{Zhang11} proposed to use Generalized Hyperbolic distributions to model volatilities to capture fat tails and skewness. \citet{BauLub02} comment on how the introduction of Student-t errors in the GARCH model may improve the fit to the data. However, the likelihood is ill-behaved as discussed in \citet{BawLub98a} and \citet{FonFerMig08}. For an illustration of this likelihood behaviour, figure \ref{contour1} presents the likelihood function for two datasets of size 150 for two parameters in the complete model we present in section 2.3. The first dataset has a well-behaved likelihood with a well-defined maximum in the true parameter values while the second dataset has an ill-behaved likelihood which goes to infinity as the parameters grow. \citet{BawLub98a} propose the use of Griddy-Gibbs sampler, which would not work in the cases where the likelihood is monotonic \citep{FonFerMig08}. Also \citet{ard08} describes a Bayesian approach to Student-t GARCH models using modified exponential prior distributions. This proposal would not work either for the situation where the likelihood is monotonic in the parameters. In this case, the choice of prior distributions may dominate the inference and posterior distributions will be similar to the prior selected. In this work, the degrees of freedom are estimated using the independent Jeffreys priors presented in \citet{FonFerMig08} which corrects the problems in the likelihood function for the Student-t model. Our proposal is a noninformative prior and does not depend on the specification of hyperparameters. This prior give the correct information regarding the curvature of likelihood functions and provide better results than the maximum likelihood estimator and informative priors. We investigate how the estimation of model parameters in the Student-t GARCH model are affected by small sample sizes, prior distributions and misspecification regarding the sampling distribution. 
 

In section 2 we present the autoregressive moving average (ARMA)  model with a GARCH component. We flexibilize the Gaussian assumption and consider Student-t error terms. We discuss the main issues related to the likelihood function and estimation of parameters such as the degree of freedom which is not usually well estimated in the literature. We present the prior considered to correct the problems with the Student-t likelihood and simulated examples which illustrate the effects of model misspecification. Section 3 presents the asymmetric GARCH model and the proposed prior distribution for the parameters of interest. The likelihood issues are discussed in the context of asymmetric models. Section 4 presents a simulation study to evaluate the performance of Bayesian estimators and Bayesian model selection. An application to the Dow Jones returns is presented in Section 5. Section 6 concludes the work with main results and future developments of the proposed models.

\section{ARMA-GARCH-M models}

Consider a univariate time series $y_t$ indexed in discrete time $t\in N_+$. For the mean term we assume an autoregressive moving average (ARMA) model and add a heteroskedasticity term (M). 
\begin{equation}\label{obsEq1}
y_t=\mu+\sum_{j=1}^{p}\phi_jy_{t-j}+\sum_{j=1}^{q}\theta_j u_{t-j}+\delta\sqrt{h_t}+u_{t},
\end{equation}
where $u_t$ are error terms with variance $h_t$ which are often modeled as Gaussian distributed, $\phi_1,\ldots , \phi_p$ are autoregression coefficcients, $\theta_1,\ldots , \theta_q$ are moving average coefficcients and $\delta$ is a parameter allowing for a direct effect of the variance term in the mean. This model is denoted as ARMA(p,q)-M(1). As follows we consider Student-t error terms and exploit the mixture model representation which could be used to capture extreme observations in financial and economic time series as follows.

\subsection{Student-t innovations with unknown degrees of freedom}

Define the error term $u_t$ as a function of a white noise $\epsilon_t$ and a positive mixing random variable $\omega_t$ as follows. 
\begin{equation}\label{eqUt}
u_t = \epsilon_t\left (\frac{\nu-2}{\nu}h_t \omega_t\right )^{1/2},\; t=p+1,\ldots,N,
\end{equation}
where $\epsilon_t\sim N(0,1)$, $\omega_t\sim IG(\nu/2,\nu/2)$. Here $N(\mu,\sigma^2)$ denotes the Gaussian distribution with mean $\mu$ and variance $\sigma^2$ and $IG(a,b)$ denotes the Inverse Gamma distribution with mean $a/b$ and variance $a/b^2$. The parameter $\nu\in \Re_+$ is responsible for the heavy tail of the sampling distribution and is considered to be an unknown constant. The mixture representation is obtained by considering the observation equation (\ref{obsEq1}) and the error term specification (\ref{eqUt}). Notice that as $\nu\to\infty$ then $u_t\sim N(0,h_t)$, while for finite $\nu$ the distribution of $u_t$ will be Student-t with $\nu$ degrees of freedom. In the mixture representation, $\omega_t$ is responsible for inflating the variance $h_t$. This is an important modeling tool in the identification of periods of larger volatility in the series. The marginal density of $u_t$ is the Student-t model given by
\begin{equation}
f(u_t;\nu,h_t)=\frac{\Gamma((\nu+1)/2)}{\Gamma(\nu/2)}((\nu-2)h_t)^{-1/2}\left (1+\frac{u_t^2}{(\nu-2)h_t}\right )^{-(\nu+1)/2}.
\end{equation}

\subsection{GARCH specification}

The variances $h_t$ are considered to be heterocedastic and given by the GARCH model which we denote by ${\rm GARCH}(r,s)$.
\begin{equation}\label{varEq2}
h_t=\omega+\sum_{j=1}^r\beta_jh_{t-j}+\sum_{j=1}^s\alpha_ju_{t-j}^2,
\end{equation}
with the restriction $\sum_{j=1}^r\alpha_j+\sum_{j=1}^s\beta_j<1$, $\omega>0$, $\alpha_i,\beta_j\geq 0$. Without loss of generality consider $\mu=0$. Consider observations $\mathbf{y_0}=(y_1,\ldots, y_{p})'$ as known and set $u_p=u_{p-1}=\ldots=u_{p-q}=u_{p-q+1}=0$. From equation (\ref{obsEq1}) we may define $u_t$, $t=p+1,\ldots,N$ recursively by

\begin{equation}
u_{t}=y_{t}-\sum_{j=1}^{p}\phi_jy_{t-j}-\sum_{j=1}^{q}\theta_j u_{t-j}-\delta \sqrt{h_{t}}.
\end{equation}

From an inferential point of view, it is convinient to rewrite Equation (\ref{obsEq1}) as

\begin{equation}\label{eqVec1}
\mathbf{y}=\mathbf{X}\phi+\mathbf{A}\theta+\mathbf{\tilde{H}}\psi+\mathbf{u},
\end{equation}
where
$\mathbf{y}=(y_{p+1},\ldots,y_N)'$, $\phi=(\phi_1,\ldots,\phi_p)'$, $\mathbf{u}=(u_{p+1},\ldots,u_N)'$, $\mathbf{\theta}=(\theta_1,\ldots,\theta_q)'$, $\mathbf{\psi}=\delta \mathbf{1}$ and the matrices\\


$\mathbf{X}
=\left (
\begin{array}{cccc}
y_p & y_{p-1} & \ldots & y_1 \\
y_{p+1} & y_{p} & \ldots & y_2 \\
 \vdots & \vdots & \vdots & \vdots \\
y_{N-1} & y_{N-2} & \ldots & y_{N-p} \\
\end{array}\right )
$, $\mathbf{A}
=\left (
\begin{array}{ccc}
 u_p  & \ldots & u_{p-q+1} \\
 u_{p+1} &  \ldots & u_{p-q+2} \\
 \vdots & \vdots  & \vdots\\
 u_{N-1} & \ldots & u_{N-q}
\end{array}\right )
$, $\mathbf{\tilde{H}}
=\left (
\begin{array}{ccc}
 \sqrt{h_{p+1}} & \ldots & 0\\
 0 & \ldots & 0\\
 \vdots & \vdots & \vdots\\
 & \ldots &  \sqrt{h_{N}}\end{array}\right )
$.\\
 
In the next section, we allow for asymmetric shock effects in order to consider more realistic set-ups for finantial time series analysis. However, the likelihood function for this model is ill behaved. This issue will be discussed and a solution will be proposed in Section \ref{SecLike}. As follows we present the asymmetric Student-t GARCH model.

\subsection{Asymmetric GARCH model\label{SecAsy}}

Consider the observation equation as specified in (\ref{obsEq1}). We now modify the variance equation (\ref{varEq2}) to accommodate asymmetric shocks in the volatility modelling. The asymmetric GARCH model for the conditional variance term is given by
\begin{eqnarray}\label{eqHt2} 
h_t & = & \omega +\lambda u_{t-1}^2f_t(u_{t-1},\gamma)+\sum_{j=1}^r\beta_jh_{t-j}+\sum_{j=1}^s\alpha_ju_{t-j}^2
\end{eqnarray}
where $\omega,\lambda>0$ are unknown parameters. The functions defining asymmetric volatilities may be specified to accommodate small and large (or positive and negative) effects. Consider the function $f_t(u,\gamma)  =  1-\exp\{-\gamma u^2\}$. If $|u|$ is large, then $f_t$ will tend to 1 and $h_t$ will be affected by $\lambda$. On the contrary, if $|u|$ is small then $f_t$ will be close to 0 and $h_t$ will be less affected by $\lambda$. This reflects different effects of small and large past error terms in the variance $h_t$. Consider the logistic function $f_t(u,\gamma)=(1+\exp\{-\gamma u\})^{-1}$, if $u\to 0$ then $f_t(u)\to 1/2$.  In this scenarium the total impact of $f_t(u)$ is $\lambda/2$. If $u\to \infty$ then $f_t(u)\to 1$ and the effect in $h_t$ is $\lambda u^2\to \infty$.  On the other hand, if $u\to -\infty$ $f_t(u)\to 0$ and the impact in $h_t \in [0,\lambda u^2[$. As a result, when $f_t(u_{t-1},\gamma)=(1+e^{-\gamma u_{t-1}})^{-1}$ the asymmetric effects generated from positive and negative shocks depend also on the shock size. If the shocks are large their effect in the variance will be significantly asymmetric. On the contrary, if the sizes are small then the model will have small asymmetry. For more details on this issue see \citet{Gon98} and \citet{Hag97}. As follows we consider two parametrisations to describe asymmetry.
\begin{enumerate}
\item If $f_t(u_{t-1},\gamma)=(1-e^{-\gamma u_{t-1}^2})$ then $f'_t(u_{t-1},\gamma)=u_{t-1}^2e^{-\gamma u_{t-1}^2}$ which allows for small and large effects.

\item If $f_t(u_{t-1},\gamma)=(1+e^{-\gamma u_{t-1}})^{-1}$ then $f'_t(u_{t-1},\gamma)=u_{t-1}e^{-\gamma u_{t-1}}(1+e^{-\gamma u_{t-1}})^{-2}$, which allows for positive and negative effects.
\end{enumerate}


\section{Likelihood issues and Bayesian Inference\label{SecLike}}

If we consider latent variables $\omega_t\sim GI(\nu/2,\nu/2)$ and the mixture representation  (\ref{eqUt}) then the likelihood function for model (\ref{obsEq1}) is given by
\begin{equation}
L(\phi,\mathbf{\theta},\psi,\nu\mid \mathbf{X},\mathbf{A},\tilde \mathbf{H},\mathbf{y},\mathbf{y_0})\propto |H|^{-1/2} \exp\left \{ -\frac{1}{2}z'H^{-1}z \right \}.
\end{equation}
with $z=\mathbf{y}-X\phi-A\theta-\tilde{H}\psi$, $H=diag(\omega_{p+1}\frac{\nu-2}{\nu}h_{p+1},\ldots,\omega_N\frac{\nu-2}{\nu}h_N)$. The estimation of the degree of freedom parameter $\nu$ is not straighforward. As discussed in \citet{FonFerMig08} the likelihood function is ill behaved and the use of naive noninformative priors such as the uniform may lead to improper posterior distributions for the parameters of interest. As showed in the paper, there is a positive probability that the maximum likelihood estimator does not exist for some data sets. This is not an issue related to the frequentist approach but an intrinsic problem of the likelihood. The following test may be applied to a given data set in order to test whether the likelihood of $\nu$ is well behaved or does not have a maximum. If the following condition is satisfied, the likelihood is not well behaved.
\begin{equation}
\sum_{i=1}^{n}(\hat z^2 - 1)^2 < 2n,
\end{equation}
where $n$ is the sample size and $\hat z_i$ are standardized residuals under normality. In this paper, we consider the correction for the likelihood proposed by \citet{FonFerMig08}, that is, we consider Jeffreys prior for the degree of freedom $\nu$ in the ARMA-GARCH set up. For the parameters $\Phi,\theta,\psi,\alpha,\beta$ we consider flat prior distribution given by
\begin{equation}
p(\Phi,\theta,\psi,\alpha,\beta\mid \nu)\propto k.
\end{equation}
While for $\nu$ we consider the independent Jeffreys prior given by 
\begin{equation}\label{nuprior}
p(\nu)\propto \left ( \frac{\nu}{\nu+3}\right )^{1/2}\left \{ \vartheta'(\nu/2) -\vartheta'((\nu+1)/2)-\frac{2(\nu+3)}{\nu(\nu+1)}\right \}^{1/2},
\end{equation}
where $\vartheta(a)=d\;log\Gamma(a)/da$ and $\vartheta'(a)=d\{\vartheta(a)\}/da$ are the digamma and trigamma functions, respectively.

\citet{FonFerMig08} present a simulated study showing that the posterior median is a better estimator for $\nu$ than the maximum likelihood estimator. They also prove that the marginal posterior distribution of $\nu$ is proper. In the frequentist context this may be seen as a problem of bias reduction of maximum likelihood estimates as presented in \citet{firth93}. Thus the likelihood are penalized by a function, the Jeffreys invariant prior, which is responsible for correcting the estimation problems. 

Other authors also reported results related to the Student-t likelihood being ill behaved such as Bauwens and Lubrano (1998). Their theorem 1 is about the flatness of the likelihood function for the Student-t model in the GARCH model. However, the suggested priors for the degrees of freedom would not work in cases where the likelihood has not a maximum as the posterior will have the same behaviour as the prior distribution. Thus sufficient prior information would be needed to well estimate this parameter. Thus, this solution will not work in situations where there is no prior information. The authors propose a Griddy-Gibbs sampler to solve the inferential problem. The method will find an estimate for the degrees of freedom, however, this will tend to the limit of the grid whenever the likelihood is too much ill behaved. \citet{Walker14} recently proposed a reference prior distribution for the degrees of freedom parameter. However, their proposal considered a discrete set of possible values for the degrees of freedom and truncated the upper limit of this set so that the Gaussian case is no considered.

The issue of likelihood functions tending to a constant will always come up whenever there are limiting cases or distributions in the model. Indeed, this is the case of the asymmetric model used here for the volatility. Regarding the asymmetric model (\ref{eqHt2}), consider the case where $\gamma\to \infty$. Then $lim_{\gamma\to \infty}f_t(u,\gamma)=1$, which implies that the likelihood function for the asymmetric GARCH model tends to a constant given by the likelihood function of the symmetric GARCH model. This condition may lead to ill behaved likelihood functions. That is, there is a positive probability that the likelihood function is a increasing function of $\gamma$. Theorem 1 of \citet{Lub98} states this problem of the likelihood. In other words, the likelihood for the asymmetric model is also ill behaved, and may be increasing in $\gamma$. Thus, the prior distribution of $\gamma$ needs to have tails that go fast enought to 0, allowing the posterior distribution to be an integrable function of $\gamma$.  An alternative is to consider the prior proposed by \citet{Lub98} which is given by
\begin{equation}\label{Lub98prior}
\pi(\gamma)=(1+(\gamma-\gamma_0)^2)^{-1},
\end{equation}
$\gamma,\gamma_0>0$. Notice that for this proposal it is required to specify a hyperparameter $\gamma_0$. The focus of this paper is the estimation of the degrees of freedom in the Student-t model for the error term. Thus we consider a proper prior for the asymmetry parameter as suggested by \citet{Lub98} but an actual derivation of Jeffreys noninformative priors for this case will be considered in a future work.

Having defined all the required prior distributions, posterior analysis may be obtained by Gibbs sampling algorithm generating from the complete conditional distributions with Metropolis steps for $\mathbf{\alpha}$, $\mathbf{\beta}$ and $\nu$. The complete conditional distributions are presented in the Appendix.

\section{Bayesian model selection\label{SecBayesTest}}

In this work model uncertainty in the sampling distribution of the data is considered through Bayes Factor computations and Bayesian hypothesis testing. As follows the main tools used are described. Consider $M$ competing models, so that we have $M$ likelihood and prior distributions, denoted respectively by $p_m(x|\theta_m) $ and $p_m(\theta_m)$, $\theta_m \in \Theta_m, m=1,\cdots, M$.
Let us introduce a discrete prior distribution over the set of $M$ alternative  models, denoted by $p(m)=\pi_m, with \sum_1^M \pi_m=1$. The joint distribution of $(x,m,\theta_m)$ is given by
$
p(x,m,\theta_m) = p_m(x|\theta_m)p_m(\theta_m) p(m).
$
Using Bayes theorem we immediately obtain the posterior distribution
$p(m, \theta_m|x)$ and  the marginal distribution $p(m|x)$ which encapsulates the uncertainty about the unknowns $(m,\theta_m)$ after observing $x$. The posterior inference in the presence of uncertainties about the correct model involves the evaluation of the posterior $p(m|x), m=1, \cdots M$ which depends on the marginal distribution $p_m(x)$ and the evaluation of $p_m(\theta_m|x)$ the posterior distribution of $\theta_m$. Let the Bayes Factor \citep{Kass95} of model 1 with respect to model 2 be
  \begin{equation}
  B_{12} =   \frac{ p(x|m_1)}{ p(x|m_2) }.
  \end{equation}
In the model choice problem one may consider the following benchmarks to decide between models. The guideline provided in \citet{Kass95} for interpretation of the Bayes factor is presented in Table \ref{TabBayesFactor}.

Let us now consider the special case of just two alternative models. A decision problem is completely specified by the triple $\{\textsl{A}, \Theta, \textsl{X}\}$, where $\textsl{A}$ is the decision space, $\Theta$ the parameter space and $\textsl{X}$ is the sample space. Let $\textsl{L}(\theta , \textsl{a})$ be a loss function for decision $a\in A$ and $\theta\in \Theta$. 

Model $m_1$ is defined by $\theta \in \Theta_1$ and the alternative model $m_2$ by $\theta \in \Theta_2$, which are denoted by $H_i, \ i=1,2$. The parameter space is partitioned in two disjoint components  $\Theta_1$ and $\Theta_2$.
 The action space is defined by two components, $\textsl{A}=\{a_1, a_2\}$, meaning that $a_i$ - the hypothesis $H_i$ is the true one and so must be accepted. Often $L(\theta, a_i) = 0 $ if $\theta \in \Theta_i$ and $K_i$ if $\theta \in \Theta_j, j\ne i$. Actually a hypothesis test is a decision rule defined on the sample space and assuming values $\{a_1, a_2 \}$, that is: $\delta: \textsl{X} \rightarrow \{0,1\}$.  As is well known from the decision making literature $a_1 \succ a_2$ if and only if $E[L(\theta,a_1) < E[L(\theta,a_2)$, where the expectation is  with respect to the posterior distributions. This is equivalent to accept $H_1$ if and only if $ \frac{P(H_1|x)}{P(H_2|x} > \frac{k_2}{k_1}$, which is equivalent to 

\begin{equation}\label{BayesFactor}
B_{12}>\frac{k_2 P(H_2)}{k_1 P(H_1)}.
\end{equation}
This will be used in the paper to chose between models of interest (e.g. Gaussian versus Student-t, asymmetric versus symmetric). Some drawbacks with the Bayesian hypothesis testing include the treatment of precise hypothesis or point null and one side hypothesis. Also the choice of the prior distribution are influential in the final results.

The computation of the predictive distribution $p(x\mid m_i)$ is not straighfoward. It is needed to consider the samples obtained from the MCMC algorithm in order to numericaly compute the precidtive distribution for each model of interest. For a given model $m_i$ \citet{NewR94} proposed the following estimator based on samples from the posterior distribution.
\begin{equation}
\hat p_1(x)=\frac{\frac{d m}{1-d}+\sum_{i=1}^{m}p(x|\theta^{(i)})\{d\hat p_1(x)+(1-d)p(z|\theta^{(i)})\}^{-1}}{\frac{d m}{(1-d)\hat p_1(x)}+\sum_{i=1}^{m}\{d\hat p_1(x)+(1-d)p(x|\theta^{(i)})\}^{-1}},
\end{equation}
where $\theta^{(1)},\ldots,\theta^{(m)}$ are generated from the posterior distribution $p(\theta|x)$. This estimator performs well for $d$ as small as $0.01$. An alternative proposal is the shifted gamma estimator proposed by \citet{Raft07}. In this proposal, the outputs of the MCMC algorithm is used to calculate a sequence of loglikelihood values $\{l_k:k=1,\ldots,n\}$ and the posterior distribution of the loglikelihoods is given by
\begin{equation}\label{ShifGa}
l_{max}-l_k\sim {\rm Gamma}(\alpha,\lambda^{-1}),
\end{equation}
where $l_{max}$ is the maximum achievable likelihood, $\alpha=d/2$, $d$ is the number of parameters in the model and $\lambda<1$. In practice, $\lambda$ is not much less than $1$. Combining the harmonic mean identity $\frac{1}{p(x)}=E\left \{\frac{1}{p(x|\theta)}\right \}$ with (\ref{ShifGa}) results in
\begin{equation}
log(p(x))=l_{max}+\alpha log(1-\lambda)
\end{equation}
In general, $l_{max}$ is not known thus $\hat l_{max}=max\{\bar l +s_l^2,l_k\}$ is used, where $\bar l +s_l^2$ is the moment estimator of $l_{max}$, $\bar l$ and $s_l^2$ are the sample mean and variance of the $l_k's$.

\vspace{1cm}

\begin{table}[htb]
\center
\caption{\label{TabBayesFactor} Guideline for interpretation of the Bayes factor $B_{12}$ which is the evidence in favour of model $M_1$ versus model $M_2$.}
\begin{tabular}{ccc}
\hline
$2ln(B_{12})$  &   $B_{12}$ & Evidence against $M_2$ \\
\hline
0 to 2 & 1 to 3 & Not worth more than a bare mention\\
2 to 6 & 3 to 20 & Positive\\
6 to 10 & 20 to 150 & Strong\\
$>$10 & $>$150 & Very strong\\
\hline
\end{tabular}
\end{table}

\section{Simulation study\label{SecSimula}}

In this section we perform a Monte Carlo simulation study to evaluate the effects of model mispecification in the results of Bayesian hypothesis testing as described in Section \ref{SecBayesTest} and also to evaluate prediction performance of different models. For each scenarium considered (sample size and sampling distribution) we simulate 100 datasets of size $T$. In this study we vary the sample size ($T=150, 500$) and the sampling distribution (Gaussian and student-t).  For each dataset several measures of model performance were computed.

Regarding the model choice problem, we define a sucess when the optimal decision selected in the Bayesian test coincides with the true model. The optimal decision is to choose the model which has Bayes Factor greater than 3 as described in table \ref{TabBayesFactor}. We compute the rate of sucess by Monte Carlo simulation. We compute Mean Squared Errors for parameters in the GARCH model for all scenarium. We also evaluate the one-step-ahead variance in order to compare the Gaussian and Student-t models. 

The model formulation considered for simulation is defined according to equations (\ref{obsEq1}) and (\ref{eqHt2}), that is,
\begin{eqnarray}\nonumber
y_t & = &\phi y_{t-1}+\theta u_{t-1}+\delta\sqrt{h_t}+u_{t}\\ \nonumber
h_t & = & \omega +\lambda u_{t-1}^2f_t(u_{t-1},\gamma)+\beta h_{t-1}+\alpha u_{t-1}^2
\end{eqnarray}
The parameters are set to be $\omega = 0.25$, $\alpha = 0.5$, $\beta = 0.1$, $\gamma = 5$, $\phi = 0.8$, $\theta = 0.1$, $\delta=0$, $\lambda=1$. The sampling distribution considered for $u_t$ is either Gaussian with $u_t=\epsilon_t h_t$ or Student-t with
\begin{equation}
u_t = \epsilon_t\left (\frac{\nu-2}{\nu}h_t \omega_t\right )^{1/2},\; t=2,\ldots,N,
\end{equation}
where $\epsilon_t\sim N(0,1)$, $\omega_t\sim IG(\nu/2,\nu/2)$.  For the degrees of freedom we considered datasets with $\nu=3$ ou $\nu=6$. 

Tables \ref{mse}, \ref{mse2} and \ref{mse3} shows the Mean Square Error (MSE) for datasets simulated (n=150,500) using the Gaussian and the Student-t models with asymmetric volatility. In the case of Gaussian data both models (Gaussian and Student-t) have similar behaviours in the estimation of all parameters ($\alpha$, $\beta$, $\lambda$, $\gamma$) as presented in table \ref{mse}. Indeed, this is a good property of Student-t models as it is able to acommodate gaussianity as a limiting case. On the other hand, for Student-t data with $3$ degrees of freedom the Gaussian model has a poor performance when compared with the Student-t model as presented in table \ref{mse2}. Notice that the Gaussian model give very high MSE for the asymmetry parameter $\lambda$. In fact, the MSE for the Gaussian model is more than 14 times larger than for the Student-t model. As expected, the same does not happen for $\nu=6$ degrees of freedom, which performs similarly to the Gaussian model as shown in table \ref{mse3}. 


\begin{table}[htb]
\begin{center}
\caption{\label{mse} Mean Square Error (MSE) for $\alpha$, $\beta$, $\lambda$ and $\gamma$ obtained in the simulated study, n=150 e n=500.}
\begin{tabular}{c|cc|cc}\hline
& \multicolumn{4}{c}{Gaussian data} \\
& \multicolumn{2}{c}{n=150} & \multicolumn{2}{c}{n=500}\\
\hline
model & Gaussian & Student-t & Gaussian & Student-t \\
\hline
$\alpha$  &  0.044 &  0.046 & 0.023 &  0.024 \\
$\beta$   &  0.003 &  0.026 & 0.001 &  0.003 \\
$\lambda$ &  0.197 &  0.386 & 0.113 &  0.142 \\
$\gamma$  & 13.006 & 13.646 & 9.718 & 10.225 \\
\hline
\end{tabular}
\end{center}
\end{table}


\begin{table}[htb]
\begin{center}
\caption{\label{mse2} Mean Square Error (MSE) for $\alpha$, $\beta$, $\lambda$ and $\gamma$ obtained in the simulated study, n=150 e n=500.}
\begin{tabular}{c|cc|cc}\hline
& \multicolumn{4}{c}{Student-t data ($\nu=3$)} \\
& \multicolumn{2}{c}{n=150} & \multicolumn{2}{c}{n=500}\\
\hline
model & Gaussian & Student-t & Gaussian & Student-t \\
\hline
$\alpha$  &  0.081 &  0.067 &  0.064 &  0.041 \\
$\beta$   &  0.007 &  0.027 &  0.008 &  0.003 \\
$\lambda$ &  1.414 &  0.739 &  5.903 &  0.403 \\
$\gamma$  & 12.319 & 13.291 & 11.193 & 12.523 \\
\hline
\end{tabular}
\end{center}
\end{table}

\clearpage
\newpage

\begin{table}[htb]
\caption{\label{mse3} Mean Square Error (MSE) for $\alpha$, $\beta$, $\lambda$ and $\gamma$ obtained in the simulated study, n=150 e n=500.}
\begin{center}
\begin{tabular}{c|cc|cc}\hline
& \multicolumn{4}{c}{Student-t data ($\nu=6$)} \\
& \multicolumn{2}{c}{n=150} & \multicolumn{2}{c}{n=500}\\
\hline
model & Gaussian & Student-t & Gaussian & Student-t \\
\hline
$\alpha$  &  0.048 &  0.051 &  0.038 &  0.037 \\
$\beta$   &  0.004 &  0.026 &  0.003 &  0.005 \\
$\lambda$ &  0.531 &  0.531 &  0.181 &  0.156 \\
$\gamma$  & 13.016 & 13.643 & 11.107 & 11.528 \\
\hline
\end{tabular}
\end{center}
\end{table}

Table \ref{prob1} presents the proportion of right decisions regarding the model which was used to simulate the data. The hypothesis testing procedure led to the correct decisions (Gaussian against Student-t model) as expected for most of scenarios. Gaussian model is selected for Gaussian data and Student-t model is selected for Student-t data. Except for small samples ($n=150$) and $\nu=6$, which is a scenario with quite light tails, in this case both models are similar. In summary, we recomend the use of the Bayes factor for model choice for the ARMA-GARCH with Student-t or Gaussian errors.

In terms of predictive variance the Gaussian model presented smaller MSE in general, with the Student-t model having a smaller MSE in scenarios with high values of $y_{T+1}$ as presented in Table \ref{mse3pred}. These scenarios are the ones with very large MSE values.

\begin{table}[htb]
\begin{center}
\caption{\label{prob1} Results for the hypothesis testing procedure based on Bayes Factors. Proportion of right decisions obtained in the simulated study, n=150 and n=500.}
\begin{tabular}{cc|cc}\hline
&& \multicolumn{2}{c}{model} \\
& & Gaussian & Student-t \\
\hline
& Gaussian & 0.86 & - \\
data (n=150) & Student-t ($\nu=3$) & - & 0.89 \\
& Student-t ($\nu=6$) & - & 0.45 \\

\hline
& Gaussian & 1.00 & - \\
data (n=500) & Student-t ($\nu=3$) & - & 0.99 \\
& Student-t ($\nu=6$) & - & 0.93 \\
\hline
\end{tabular}
\end{center}
\end{table}

\newpage
\clearpage

\begin{table}[htb]
\begin{center}
\caption{\label{mse3pred} Mean Square Error (MSE) for the predictive variance obtained in the simulated study in the one-step-ahead prediction, n=150 e n=500.}
\begin{tabular}{cc|cc|cc}\hline
& & \multicolumn{4}{c}{Model} \\
& & \multicolumn{2}{c}{n=150} & \multicolumn{2}{c}{n=500}\\
\hline
& & Gaussian & Student-t & Gaussian & Student-t \\
\hline
& Gaussian  &  7.4046 & 16.9097  & 830.23 &  791.66 \\
Data & Student-t ($\nu=3$) & 15.26  & 20.70  & 4.93  &  3.63\\
& Student-t ($\nu=6$) & 11650.43  &  11186.32  & 9.04 & 9.34 \\
\hline
\end{tabular}
\end{center}
\end{table}

\section{Application}

In this section we present an application of the proposed model to study the dynamics in the daily Dow Jones returns. Our main goal in this application is to highlight the predictive advantages of the Student-t model when compared to the Gaussian model in periods of high volatility of the series. In this context, we analysed the daily Dow Jones index of the New York stock market and is given by $r_t=100 ln(DJ_t/DJ_{t-1})$. We present results from 02/01/2007 to 31/12/2008, with the first period going up to 30/05/2008 used for estimation which results in 354 observations and the final part used for predictive performance evaluation. We sucessively update the data with one observation after prediction of this point resulting in a total of 503 observations. The period selected for prediction has two different volatility regimes. The first period from 02/06/2008 to 12/09/2008 is before the bankruptcy of Lehman Brothers Bank. This event and the Estate Market crisis resulted in large volatility in the american stock market from 15/09/2008. This instability decreased from December 2008. Thus, the data selected for prediction will allow comparison of predictive performance of the models for different kinds of volatility.

In this application we fitted three kinds of asymmetric models: the logistic and exponential as presented in subsection \ref{SecAsy} and a modified logistic model which we call logistic model 2 which is defined as follows. As pointed out by \citet{Gon98} in the finantial market good news are associated with positive shocks which tend to result in small volatility while bad news tend to result in negative shocks which will usually be associated with large volatilities. In this regard, we consider also the logistc function 2 defined by
$f_t(u)=(1+\exp\{\gamma u\})^{-1}$. Thus, we have six model formulations:  Gaussian GARCH with logistic asymmetry types 1 and 2, Gaussian GARCH with exponencial asymmetry, Student-t GARCH with logistic asymmetry types 1 and 2 and Student-t GARCH with exponencial asymmetry. In addition, we compute prediction based on a hybrid approach which considers the results from hypothesis tests as presented in Section 4. We fitted a ARMA(3,0) for the mean which resulted in a stationary time series and a GARCH(1,1) for the variance modeling. In the MCMC algorithm we considered ten thousand iteractions and the acceptance rate was tunned to be in the range 0.2 to 0.4. We follow \cite{Awar05} and considered the squared observed returns as proxy for the variance of the process in the computation of Mean Squared Errors (MSE)  for the four competing models. 

\begin{figure}[htb] 
\begin{center} 
\begin{tabular}{cc}
\includegraphics*[height=6cm,width=6cm]{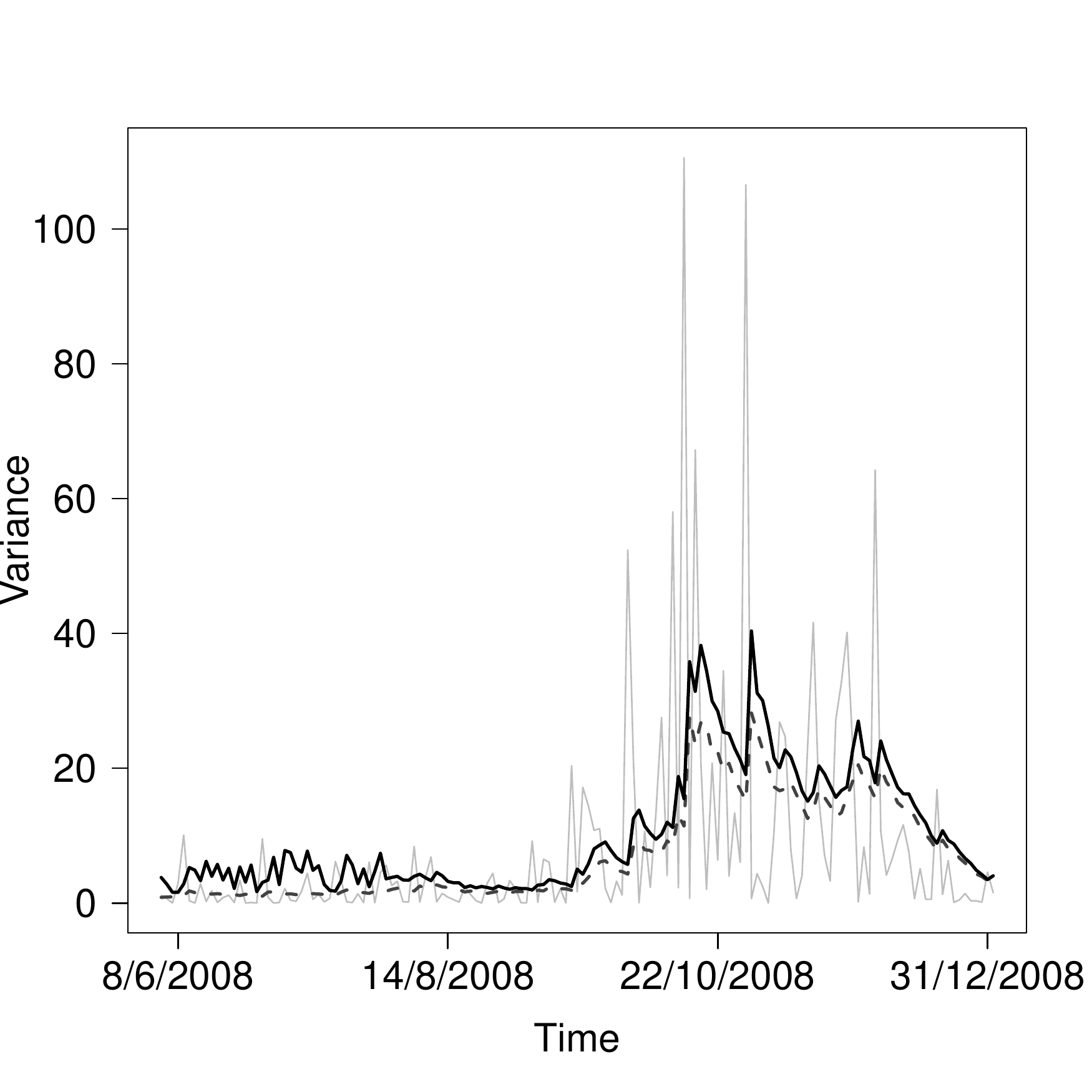} &
\includegraphics*[height=6cm,width=6cm]{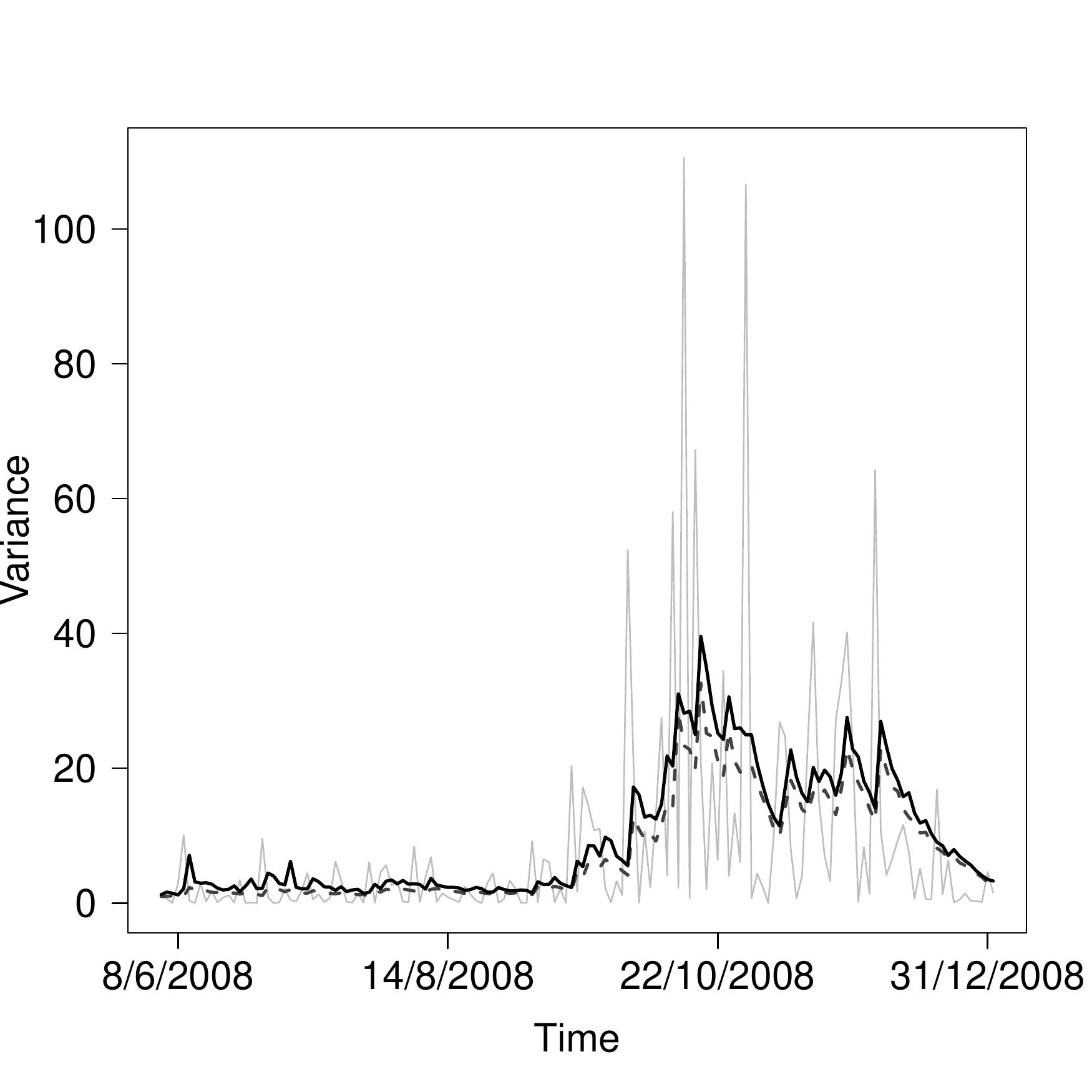}\\
(a) Logistic asymmetric model & (b) Logistic asymmetric model 2.\\
\includegraphics*[height=6cm,width=6cm]{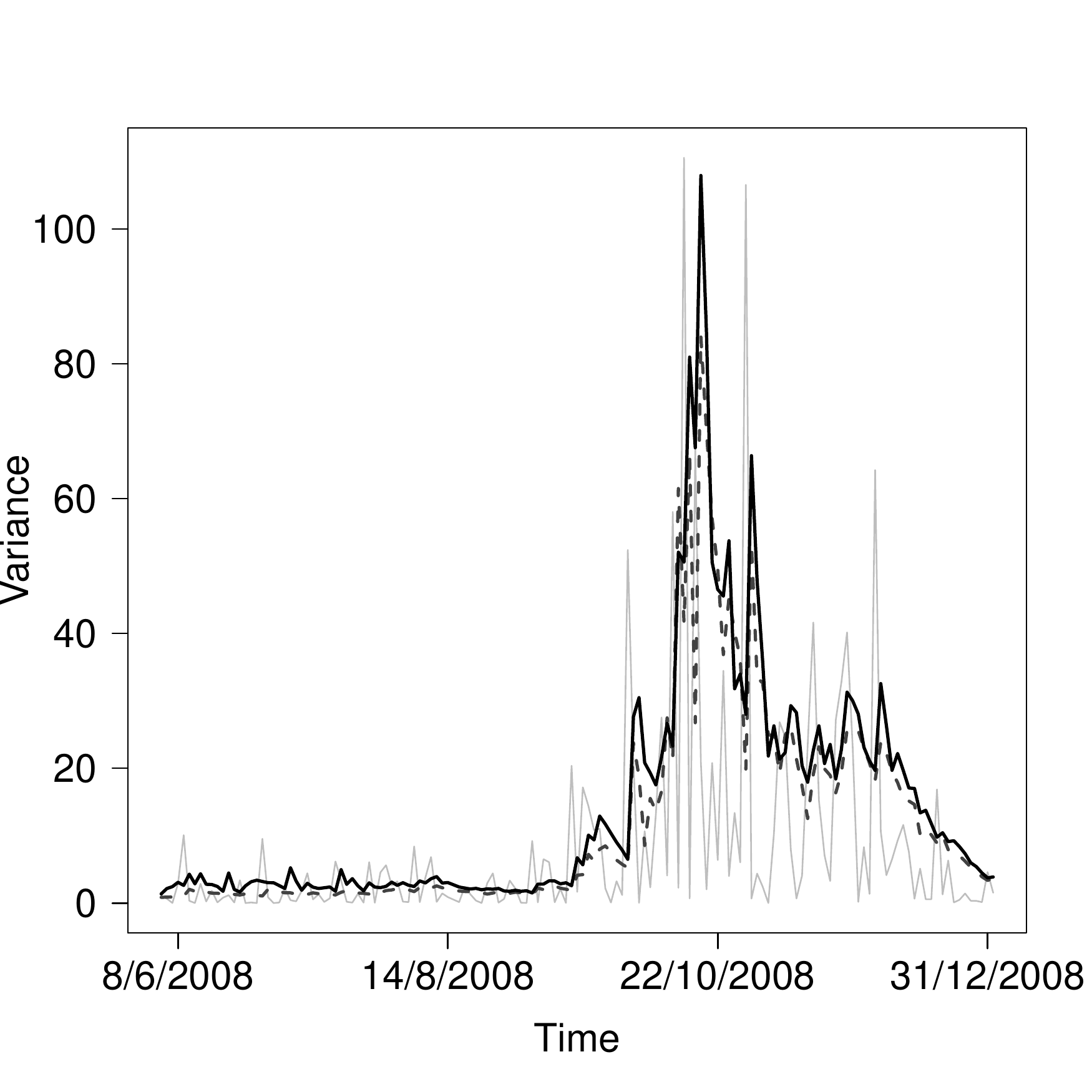} &\\
(c) Exponential asymmetric model.
\end{tabular}
\caption{ Squared returns (light grey full line), Gaussian (grey dashed line) and Student-t (black full line) predictions with asymmetric volatility model.}
\label{ret2}
\end{center}
\end{figure}

Figure \ref{ret2} present the variance proxy evolution through time and the predictions. Note that the exponencial model is able to capture the very large volatility observed from 09/2008 to 12/2008. This is confirmed by the large advantage of this model for this period using the Bayes factor as a model comparison measure as presented in table \ref{tabAp1}. 

In the context of Gausssian and Student-t models comparison, the majority of the testes indicate the Student-t model as the best fit. From 02/06/2008 to 12/09/2008 the models behave similarly and model selection does not give a large preference to the Student-t model. On the contrary, from 15/09/2008 - 31/12/2008 the Student-t model is strongly preferred when compared to the Gaussian model. This is due to the presence of large volatility in this period.

\begin{figure}[htb] 
\begin{center} 
\begin{tabular}{cc}
\includegraphics*[height=6cm,width=6cm]{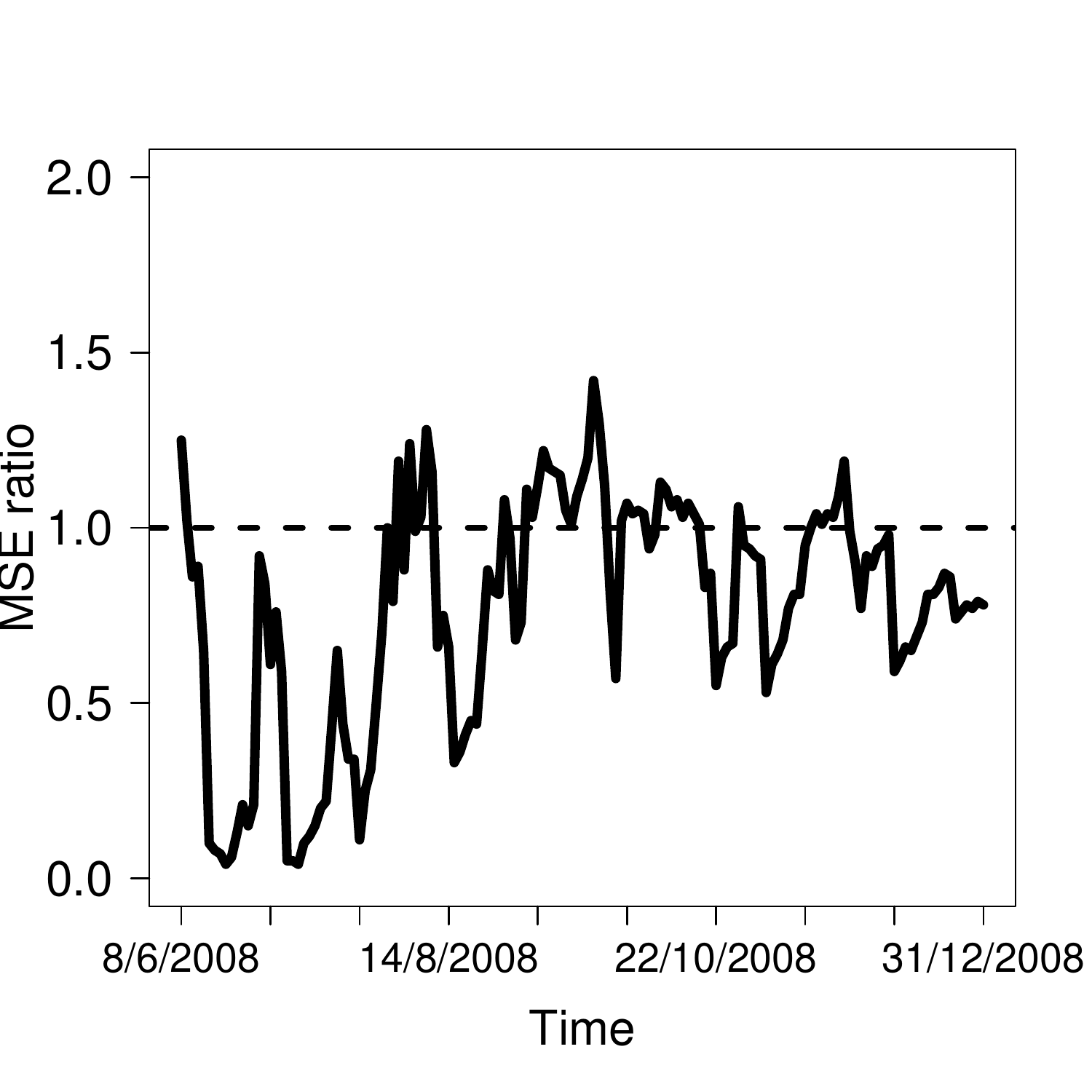} &
\includegraphics*[height=6cm,width=6cm]{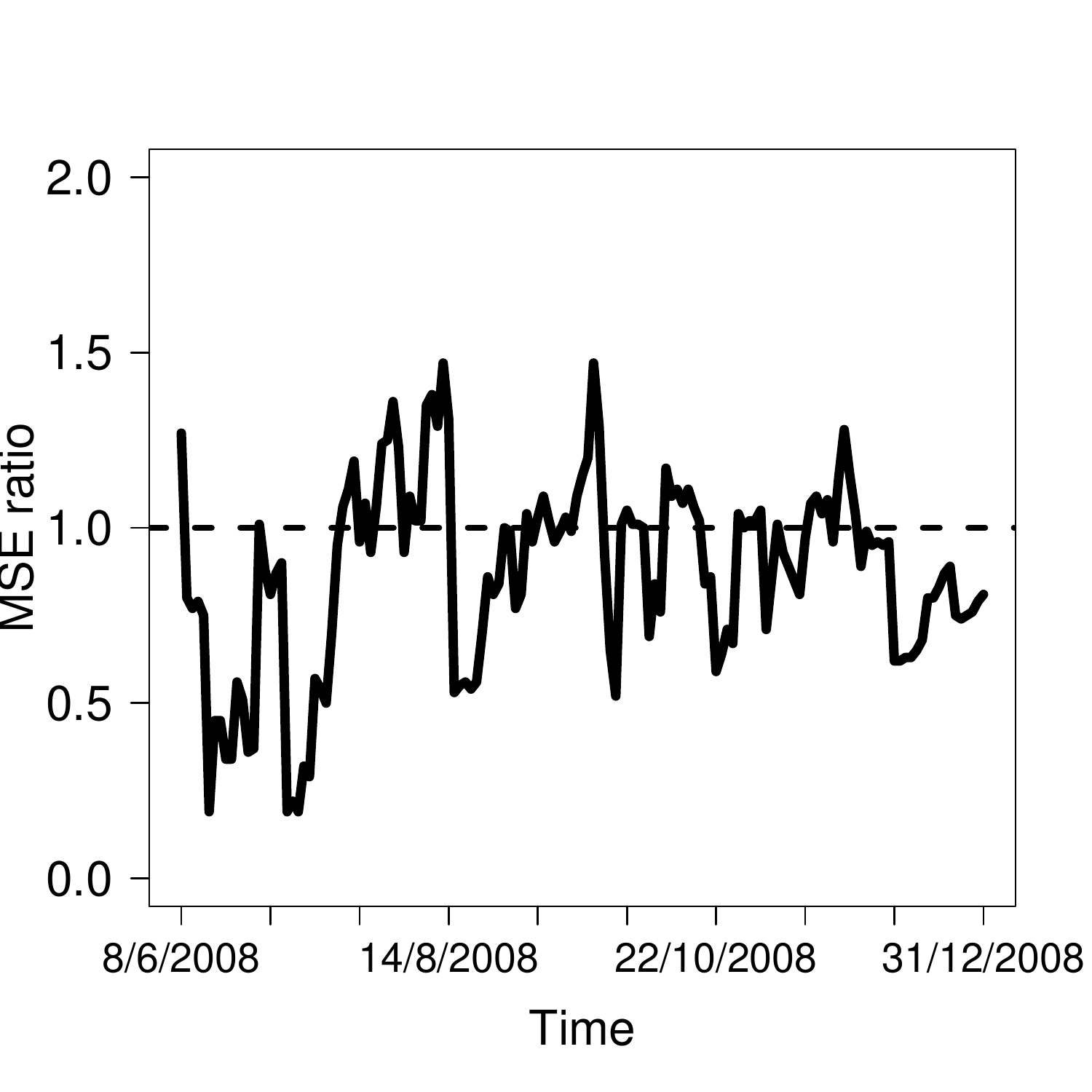}\\
(a) Logistic asymmetric model & (b) Logistic asymmetric model 2.\\
\includegraphics*[height=6cm,width=6cm]{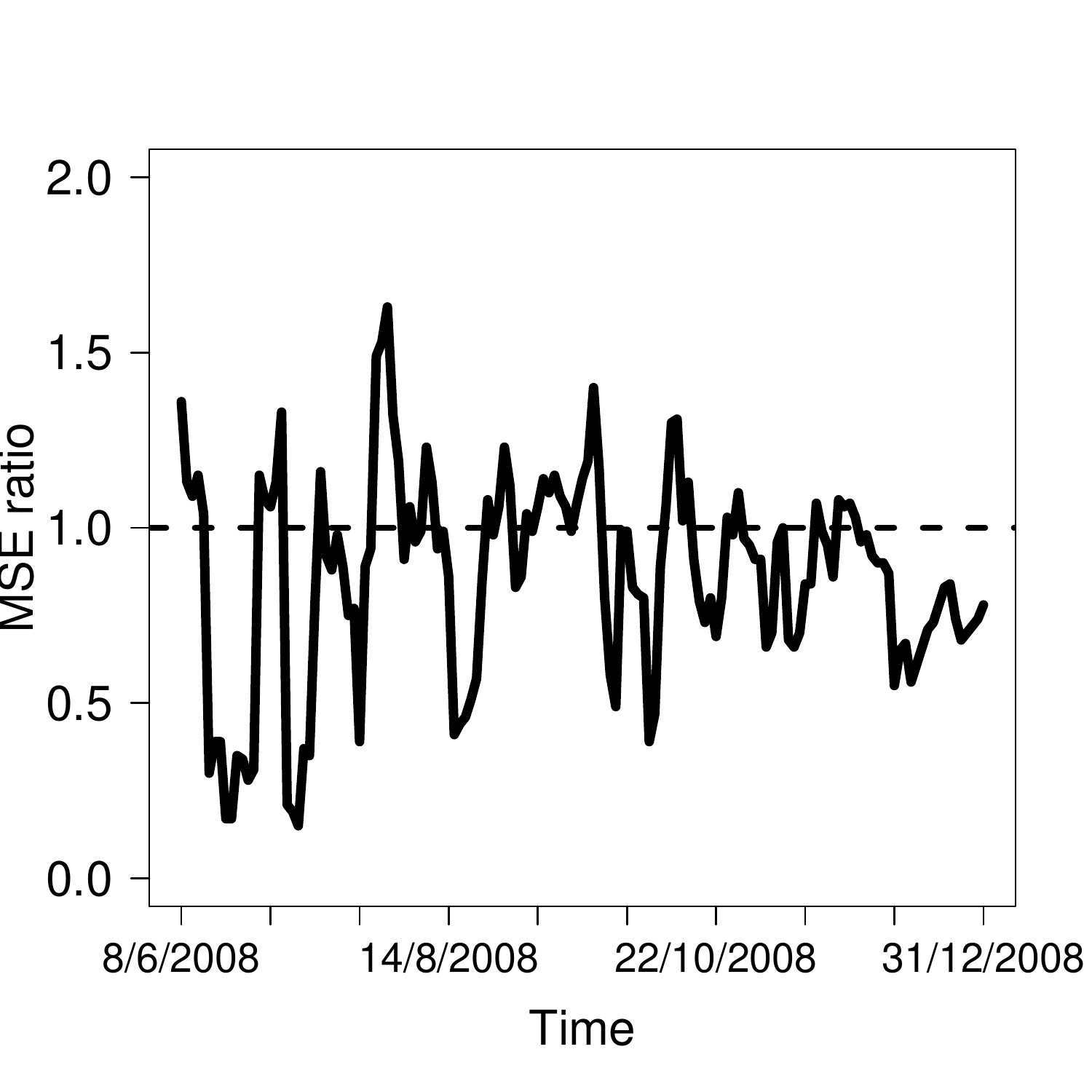} &\\
(c) Exponential asymmetric model.
\end{tabular}
\caption{Mean squared error ratio between the Gaussian and Student-t models with asymmetric volatility for the past five days. Values greater than one indicate the Student-t model is preferred.}
\label{Ratio}
\end{center}
\end{figure}

In order to analyse closely the predictive performance for the Gaussian and Student-t models in the two different regimes we define the MSE ratio for the Gaussian against Student-t model for the last five observed days given by 
\begin{equation}
I_{t+5}=\frac{\sum_{n=1}^{5}(\hat h_{t+n}^G-h_{t+n})^2}{\sum_{n=1}^{5}(\hat h_{t+n}^{ST}-h_{t+n})^2},
\end{equation}
with $h_{t+n}$ the conditional variance (squared observed returns), $\hat h_{t+n}^G$ and $\hat h_{t+n}^{ST}$ the one step ahead prediction for the conditional variance in the Gaussian and Student-T models, respectively. Figure \ref{Ratio} presents the MSE ratio (Gaussian versus Student-t) for $I_{t+5}$ across time for the logistic, logistic 2 and exponential asymmetric models, respectively. The evolution in panel (a) and (b)  indicate that for the logistic model and logistic model 2, periods of time with high volatility tend to have $I_{t+5}$ greater than 1, which suggests that the Student-t model would be more indicated for these periods. For the exponential asymmetric model (panel (c)) this difference is less evident, although there are periods in which the Student-t model would be more recommended. Indeed, the correlation between $I_{t+5}$ and the mean squared returns in the five past days is 0,54 for the logistic model 1, 0,42 for the logistic model 2 and it is 0,29 for the exponential model.

\begin{table}[htb]
\begin{center}
\caption{Model selection test for the logistic asymmetry 1, logistic asymmetric 2 and exponential. \label{tabAp1}}
\begin{tabular}{cccc}\hline
Time period & Logistic 1 & Logistic 2 & Exponential\\ \hline
02/06/2008 - 12/09/2008 & 12.32 & 9.06 & 8.41\\
15/09/2008 - 31/12/2008 & 523.27 & 449.78 & 820.75\\
02/06/2008 - 31/12/2008 &  272.94 & 233.86 & 422.76\\
\hline
\end{tabular}
\end{center}
\end{table}

\newpage
\clearpage

\section{Conclusions}


This work shows how Student-t smooth transition models can be used adequately from a Bayesian point of view. Our main interest in this paper was to investigate the prediction and estimation performances of Student-t sampling distributions based on the independent Jeffreys prior assumption for the degrees of freedom parameter in the context of smooth transition models. The degrees of freedom in the Student-t model are difficult to estimate and it was considered independent Jeffreys prior to solve this estimation problem. The likelihood tends to a constant with positive probability. This behaviour is intrisic to the likelihood and not corrected by usual parametric priors such as exponential and gamma. This is due to the existance of limiting cases (Gaussian and symmetric models) for the sampling distributions assumed for the data. Thus, we suggested the use of independent Jeffreys priors, which are proved to lead to proper posteriors and have nice frequentist properties \citep{FonFerMig08}.

The proposed prior gave positive results as presented in the simulated study. For model selection we use Bayesian Hypothesis testing based on the numerical computation of predictive distributions. The Bayesian test based on Bayes factors was effective in the decision between Gaussian and Student-t models. Furthermore, the simulated study indicate that some factors such as parameter estimation and prediction one step ahead may increase competitivity of Student-t models. For instance, this is the case when data is simulated from Student-t models with relatively heavy tails ($\nu=3$). 

For the Dow Jones application, in general the Gaussian model had the best predictive performance with the Student-t model being preferable in large volatility periods, specially for the logistic asymmetric model 2. This is crucial to correctly estimating the uncertainty in periods of large volatility in the market.

In this work the informative prior proposed by \citet{Lub98} was considered for the smooth transition parameter. However, this prior depends on hyperparameters, thus in a future research it will be considered the development of new reference priors for this problem which would not depend on hyperparameter specification.

\section*{Appendix}

This appendix presents the complete conditional distributions used in Markov Chain Monte Carlo sampling of model parameters for the Student-t GARCH model with Independent Jeffreys prior for the degrees of freedom. The code where this algorithm is implemented is available for general R package users.

\begin{itemize}
\item Parameter $\Phi$:

$$\Phi\mid \theta,\psi,\alpha,\beta, \mathbf{X},\mathbf{y},\mathbf{y_0}\sim N\left ( \hat \Phi, (X'H^{-1}X)^{-1}\right ),$$
where
$\hat\Phi=(X'H^{-1}X)^{-1}(X'H^{-1}\mathbf{y}-X'H^{-1}A\theta-X'H^{-1}\tilde{H}\psi)$.

\item Parameter $\theta$:
$$\theta\mid \Phi,\psi,\alpha,\beta,\mathbf{X},\mathbf{y},\mathbf{y_0}\sim N\left (\mu_{\theta},V_{\theta}\right),$$
where $V_{\theta}=(A'H^{-1}A)^{-1}$ and $\mu_{\theta}=(A'H^{-1}A)^{-1}(A'H^{-1}\mathbf{y}-A'H^{-1}X\Phi-A'H^{-1}\tilde{H}\psi)$.

\item Parameter $\psi$: 
$$\psi\mid \Phi,\theta,\alpha,\beta,\mathbf{X},\mathbf{y},\mathbf{y_0}\sim N\left (\mu_{\psi},V_{\psi}\right),$$
where $V_{\psi}=(\tilde{H}'H^{-1}\tilde{H})^{-1}$ and $\mu_{\psi}=(\tilde{H}'H^{-1}\tilde{H})^{-1}(\tilde{H}'H^{-1}\mathbf{y}-\tilde{H}'H^{-1}X\Phi-\tilde{H}'H^{-1}A\theta)$.

\item Parameter $\mathbf{\alpha}$ and $\mathbf{\beta}$:
$$p(\alpha,\beta\mid \ldots)=\prod_{j=1}^{N-p} \phi(y_t\mid\mu_t,h_t),$$
where $\mu_t=\sum_{j=1}^{p}\phi_jy_{t-j}+\sum_{j=1}^{q}\theta_j u_{t-j}+\sum_{j=1}^k\psi_jh_{t-j}$ and $\phi(.)$ is the Gaussian density function. We define independent proposal distributions  given by
$$\alpha_i\sim N(\hat{\alpha_i},\hat{\Sigma}_{\alpha_i}),\ \ \ \ \ \beta_i \sim N(\hat{\beta_i},\hat{\Sigma}_{\beta_i})
$$

\item Parameter $\nu$: 
\begin{eqnarray*}
p(\nu| \Phi,\theta,\psi,\alpha,\beta, \mathbf{X},\mathbf{y},\mathbf{y_0}) &\propto& |H|^{-1/2} \exp\left \{ -\frac{1}{2}z'H^{-1}z \right \} \frac{(\nu/2)^{(N-p)\nu/2}}{\Gamma(\nu/2)^{N-p}}\\ &\times &\left (\prod_{t=1}^{N-p}\omega_t\right )^{-\nu/2-1}\exp\left \{-\frac{1}{2}\sum_{t=1}^{N-p}\frac{\nu}{\omega_t}\right \} p(\nu),
\end{eqnarray*}
where $z=(\mathbf{y}-X\phi-A\theta-\tilde{H}\psi)$.
\end{itemize}

\clearpage
\newpage

\bibliographystyle{Chicago}

\newcommand{\sortnoop}[1]{}

\clearpage
\newpage

\clearpage
\newpage

\end{document}